\documentclass[12pt,a4paper]{article}
\usepackage{natbib}
\usepackage{geometry}
\usepackage[latin1]{inputenc}
\usepackage[T1]{fontenc}
\usepackage{graphicx,graphics,textcomp}
\usepackage{fancyhdr}
\usepackage{lscape,epsfig,amssymb,amsmath}
\usepackage[]{subfigure}
\usepackage{rotating,placeins}
\usepackage{lscape}
\usepackage{setspace}
\newenvironment{myindentpar}[1]%
  {\begin{list}{}%
     {\setlength{\leftmargin}{#1}}%
     \item[]%
  }
  {\end{list}}

\title{Advances in Search and Rescue at Sea}
\author{{\O}yvind Breivik
\footnote{Final version published as Breivik, {\O}, A A Allen, C Maisondieu and M Olagnon (2012). 
Advances in Search and Rescue at Sea, \emph{Ocean Dynam}, 10.1007/s10236-012-0581-1}
\thanks{Corresponding author. E-mail:
\texttt{oyvind.breivik@ecmwf.int}}
\thanks{ECMWF, Shinfield Park, Reading, RG2 9AX, United Kingdom. On leave from the
Norwegian Meteorological Institute}, 
\and Arthur A Allen\thanks{US Coast Guard, Office of Search and Rescue, New
London, CT, USA},
\and Christophe Maisondieu\thanks{IFREMER, Hydrodynamique et
Oc\'{e}ano-M\'{e}t\'{e}o, Plouzane, France},
\and Michel Olagnon\footnotemark[4]
}

\begin{document}
\maketitle                                                                                                          

\begin{abstract}
A topical collection on ``Advances in Search and Rescue at Sea'' has appeared
in recent issues of \emph{Ocean Dynamics} following the latest in a series of
workshops on ``Technologies for Search and Rescue and other Emergency Marine
Operations'' (2004, 2006, 2008 and 2011), hosted by IFREMER in Brest, France.

Here we give a brief overview of the history of search and rescue at sea
before we summarize the main results of the papers that have appeared in
the topical collection.


\end{abstract}

\section{A brief history of SAR planning}
Measuring and predicting the drift of search and rescue (SAR) objects has
come a long way since \citet{pingree44} made the first drift or ``leeway''
study of life rafts and presented it as ``Forethoughts on Rubber
Rafts''. The data were unfortunately of limited value, but the general method
differed little from that of the earliest successful leeway study by
\citet{chapline60} who estimated ``The drift of distressed small craft''
using visual observations of drift nets to establish the current while
simultaneously estimating the angle and speed with which the object drifted
relative to the wind. This method of conducting leeway studies is known as
the \emph{indirect method} as it indirectly measures the motion of the object
relative to the ambient current (the leeway).  The method reigned supreme
(eg, \citealt{hufford76}) until the 1990s with the possible exception of
\citet{suzuki77} who attempted to log the motion relative to the ambient
current using a bamboo pole partly submerged and attached to the side of
the ship by string. It should be obvious that the precision of these early
experiments was not impressive, but the results were still of remarkable
importance in the everyday work of rescue centres around the world.

In 1944 The United States Navy Hydrographic Office issued a manual on ``Methods
for locating survivors adrift at sea on rubber rafts'' \citep{usnho44} which
summarized much of the current knowledge at the time of how objects on the
sea surface would drift and how to conduct the search.  The mathematical
field of search theory and the wider topic of operations research (OR) grew
out of a need to respond to the German submarine threat during the second
world war. The early work was pioneered by Koopman, who after having provided
a working manual \citep{koopman46} of search and screening outlined the
fundamentals of search theory in a seminal series of papers 
\citep{koopman56a,koopman56b,koopman57}. Without a theory of search the field of search and rescue
would not exist and without a theory of how the object moves, there is no
way to define the search area for a moving target \citep{washburn80}, so
the two fields of object drift and search theory grew up together in the
post-war years.  We refer to the combined effort of modelling the object
drift and optimally allocating the search effort as SAR planning.  In the
1950s the United States Coast Guard (USCG) first applied the principles
of search theory to SAR planning when it published its search planning
doctrine in a SAR manual. Since computers were not widely available, the
methods were simplified and adapted for manual calculation.  Around 1970 the
USCG implemented the first computer-based search and rescue planning system
(SARP) which was a computer implementation of the manual methods in the SAR
manual. In 1974 the USCG implemented the first Bayesian SAR planning system,
the Computer Assisted Search Planning (CASP), see \citet{richardson80}.
CASP was among the first applications of computer-assisted Bayesian methods
(See \citealt{mcgrayne11} for a popular account of the post-war applications
of Bayesian methods in search theory and \citealt{koopman80} for a
comprehensive account of its early history).  For more details on search theory,
see \citet{stone89,fro01} and the upcoming encyclopedic entry
by \citet{stone13}.

CASP produced probability distributions by Monte Carlo methods, generating an
ensemble of particle trajectories to estimate the location of the search object
as a function of time.  The trajectories accounted for the uncertainty of the
initial position of the search object and moved the particles in accordance
with a primitive drift model. This model relied on historical
ship recordings of surface currents on a $1^\circ \times 1^\circ$ monthly
climatology grid and wind fields from the US Navy Fleet Numerical Oceanography
Center (FNOC) on a $5^\circ \times 5^\circ$ grid at 12 hour intervals forecast
to 36 hours into the future.  After an unsuccessful search, CASP computed
the Bayesian posterior distribution for the location of the search object
at the time of the next search by accounting for unsuccessful search and
motion due to drift.  A less coarse $3^\circ \times 3^\circ$ resolution ocean
model without tides was added in 1985.  There were several evaluations of
SARP and CASP drift estimates using satellite tracked buoys during the early
1980s \citep{murphy85}.  Both SARP and CASP had mixed records at predicting the drift of
search objects and very limited capabilities on or inside the continental
shelf due to the coarse forcing fields.

Near-real time surface current measurements near the last known position
(LKP) are essential to SAR operations. The USCG devised the self-locating
datum marker buoy (SLDMBs) based on the Code-Davis drifters developed
in the 1980s \citep{dav85a}. As Argos transmitters became smaller and
global positioning system (GPS) receivers more reliable and affordable this
eventually led to operational use of SLDMBs in SAR operations \citep{allen96}.
When air deployment of SLDMBs was approved in January 2002 their use became
standard routine with most SAR cases, representing a major advancement in
the real-time acquisition of surface currents. They remain an essential
tool for rapidly establishing the currents near the presumed point of
the incident. A new generation of commercially available light-weight
GPS-based SLDMBs that can be deployed from aircraft (adhering to the NATO
A-size sonobuoy standard dimensions) is now appearing. These new drifters
have a much higher report frequency as they rely on the Iridium satellite
network rather than ARGOS. The new generation SLDMBs will also open up new
possibilities for physical oceanographers as the cost has come down while
precision and reliability have improved greatly compared with earlier models.

With the advent of high-resolution operational ocean models and the continued
improvement of numerical weather prediction models (NWP), the potential
for making more detailed predictions of the fate of drifting objects grew
in the 1990s, and although the improved weather forecasts led to better
forcing, drift models remained somewhat impervious to the advances in
ocean modelling and numerical weather forecasting. This can perhaps best be
understood in light of the great uncertainties in the drift properties of
SAR objects. Without a proper estimate of the basic drift properties and
their associated uncertainties, forecasting the drift and expansion of a
search area remains difficult. An important change came when the \emph{direct
method} for measuring the leeway of a drifting object became common practice
\citep{all99,all05,bre11,hod98}. The direct method measures the object's
motion relative to the ambient water using a current meter.  Current meters
small enough and flexible enough to be towed or attached directly to a SAR
object started to become available in the 1980s, and since then almost all
field experiments on SAR objects have employed a direct measurement technique
\citep{all99,bre11,maisondieu10}. The direct method, together with a rigorous
definition of \emph{leeway} as
\begin{myindentpar}{1.5cm}
  Leeway is the motion of the object induced by wind (10~m reference
  height) and waves relative to the ambient current (between 0.3
  and 1.0~m depth)
\end{myindentpar}
and finally the decomposition of leeway coefficients in \emph{downwind} and
\emph{crosswind} components makes it possible to follow a rigorous procedure
for conducting leeway field experiments.  See \citet{all99,bre08,bre11}
for further details.

It was not until the 2000s that all the necessary components required for fully
stochastic modelling using high-quality drift coefficients and detailed current
and wind forecasts were in place. The first operational leeway model to employ
the USCG table of drift coefficients \citep{all99} with high-resolution ocean
model current fields and near-surface wind fields went operational in 2001
(see \citealt{hac06,bre08,dav09}).

The modern era of \emph{SAR planning} involving the Bayesian posterior
updates after the search began in 2007 when USCG launched the Search And
Rescue Optimal Planning System (SAROPS), see \citet{kratzke10}. SAROPS employs
an environmental data server that obtains wind and current predictions from
a number of sources.  It recommends search paths for multiple search units
that maximize the increase in probability of detection from an increment of
search. As with CASP, it computes Bayesian posterior distributions on object
location accounting for unsuccessful search and object motion.

By the late 2000s it was clear that although the level of sophistication and
detail had grown dramatically since the early days of drift nets and CASP the
uncertainties in SAR predictions remained stubbornly high. The fundamental
challenge of estimating and forecasting search areas in the presence of
large uncertainties remains essentially the same, even though certain error
sources have been diminished. The slow progress that has been made over the
past decades in reducing the rate of expansion of search areas (perhaps the
single best estimate of improvement) is an unavoidable consequence of SAR
planning being at ``the top of the food chain'' in the sense that errors
creep in from the current fields, the wind fields, missing processes (e.g.,
wave effects, see \citealt{bre08,rohrs12}), the last known position and
not least from poor estimates of the real drift properties of the object.
Indeed, sometimes the type of object may not even be known, effectively
making the modelling exercise into an ensemble integration spanning a range of
object categories. All these error sources accumulate and make SAR planning
as much art as science, where rescuers still often rely as much on their
``hunches'' as on the output of sophisticated prediction tools. The fact
that the majority of SAR cases occur near the shoreline and in partially
sheltered waters \citep{bre08} compounds the difficulties as the resolution
of operational ocean models in many places of the world is still insufficient
to resolve nearshore features.

\section{The state of the art of drift prediction}
Throughout the last decade these advances and obstacles to further progress
have been presented mainly through a series of workshops organized on
``Technologies for Search and Rescue and other Emergency Marine Operations''
(2004, 2006, 2008 and 2011, see \citealt{bre05}) organized by the French marine
research institute (IFREMER) with support from the Norwegian Meteorological
Institute, USCG, the French-Norwegian Foundation and the Joint WMO-IOC
Technical Commission for Oceanography and Marine Meteorology (JCOMM).  As the
last of these workshops drew near we decided that it was time to put some of
the advances on a more academic footing by publishing a special issue, and
\emph{Ocean Dynamics} agreed to arrange a topical collection on ``Advances
in search and rescue at sea''.  This topical collection focusses on recent
advances in the understanding of the various processes and uncertainties
that have a bearing on the evolution of trajectories at the sea surface,
from the drift properties of the objects themselves to the quality of the
forcing fields.

The diffusivity of the ocean is an important factor when reconstructing
the dispersion of particles either based on observed or modelled vector
fields. In either case the dispersion is to the lowest order governed by the
advection-diffusion equation \citep{tay21} by assuming an ``eddy-diffusivity''
coefficient. In many cases this simple stochastic model is sufficient
for estimating the dispersion of SAR objects over relatively short
time periods.  \citet{dominicis12} report carefully evaluated estimates
of the eddy diffusivity from a large data set of drifter trajectories in
the Mediterranean Sea. Such regional (and possibly seasonal) estimates of
diffusivity and the integral time scale should be carefully considered as
their impact on the dispersion of SAR objects may be substantial.

Stochastic ensemble trajectory models of drifting objects normally
employ deterministic (single-model) current and wind vector fields
and perturb the trajectories either with a random walk diffusivity
\citep{bre08,dominicis12} or with a more sophisticated second-order random
flight model \citep{spaulding06,gri96,ber02}. However, the advent of true
ocean model ensembles \citep{bertino08} have now opened up the possibility of
exploiting a full vector field ensemble for estimating drift and dispersion
in the ocean. \citet{melsom12} compared the dispersion of passive tracers in
a 100-member ensemble of the TOPAZ ocean prediction system to the dispersion
found adding random flight perturbations to the ensemble mean vector field
and a deterministic vector field. The results are not conclusive in favour of
the full ensemble, which is important to keep in mind when considering the
cost-benefit of such computationally expensive operational ocean forecast
systems.  An alternative to a full model ensemble is to employ multi-model
ensembles (see \citealt{rix07,rix08,van09}), which is what \citet{scott12}
did when they assembled five model reanalyses and compared the weighted
average with observed trajectories in the equatorial Atlantic.

Several workers \citep{bar12,kohut12,frolov12,kuang12,abascal12} investigated
the potential for high-frequency (HF) radar monitoring systems to supply near
real-time current fields to reconstruct the trajectories and the dispersion
of drifting objects in the coastal zone.  \citet{kohut12} explored the impact
on search areas from switching to an optimal interpolation (OI) scheme for
calculating total vectors from radial vector fields. Such techniques for
extending the range of HF radars (see also \citealt{bar12} discussed below)
can make a significant difference when investigating nearshore SAR cases.

HF radar fields and drifter studies can be used to evaluate the quality
of ocean model current fields. Since the rate of expansion of search areas
depends intimately on the quality of the forcing, it remains very important
to establish good error estimates for each ocean model being used for SAR
prediction. \citet{kuang12} assessed the New York Harbor Observing and
Prediction System (NYHOPS) using boths SLDMBs and HF currents. They found
good agreement between model, HF radar and three drifter trajectories in
the Middle Atlantic Bight and were able to quantify the root-mean-square
differences between the modelled NYHOPS and the observed HF fields.

HF short-term prediction of surface current vectors out to typically 12-24
hours is a technique with great potential for near-shore SAR operations.
\citet{bar12} employed open modal analysis (OMA, see \citealt{lekien04}) to
decompose the vector field into divergent and rotational modes within the HF
domain along the complex coastline of northern Norway (see \citealt{whelan10}
for a description of the radar deployment). They then predicted the short-term
variation of the amplitudes of the most energetic modes based on a relatively
short history of archived vector fields, giving short-term forecasts out
to 24 hours.  \citet{frolov12} chose empirical orthogonal functions (EOFs)
instead of normal modes and then employed an autoregressive method to make
short-term predictions out to 48 hours for an HF network in Monterey Bay.

Although the direct leeway field method was established as the superior
technique for establishing the leeway of drifting objects already in the
late 1980s, the technique was only recently presented in the open literature
by \citet{bre11}. \citet{bre12b} explored how the technique can be applied
to relatively large objects such as shipping containers and combined the
field results with estimates from earlier work on shipping containers by
\citet{dan02} to estimate how the drift varies with immersion.

Most trajectory models for small surface objects ignore the direct wave
excitation and damping since only waves whose wave length is comparable to
the dimensions of the object will excert a significant force on the object
\citep{bre08,mei89}. Since SAR objects are typically smaller than 30~m their
resonant ocean waves will have only negligible energy. However, waves will
also affect an object through the Stokes drift \citep{phi77,holthuijsen07},
which is a Lagrangian effect not visible in an Eulerian frame of
reference. \citet{rohrs12} explored how the Stokes drift affects surface
drifters with and without leeway directly and through the addition of the
Coriolis-Stokes effect to the momentum equation. The term adds an additional
deflection to upper-ocean currents caused by the Coriolis effect acting on
the Stokes drift. This has clear relevance for the operational forecasting
of SAR objects as well as for the interpretation of SLDMB trajectories,
although it is not clear yet how large the effect is for real-world search
objects that also move under the direct influence of the wind.

Finally, the importance of being able to estimate the point of an accident
based on a debris field was made poignantly clear after the AF447 aircraft
accident on 1 June 2009 in the equatorial Atlantic (see \citealt{stone11}
for an account of the search effort following the accident). Using SAR
trajectory models for backtracking is not trivial since it effectively means
reversing the (usually weakly nonlinear) processes that propel the object.
In principle it is better to run a model forward and iterate, as \citet{bre12}
has demonstrated, but nevertheless direct backtracking can be employed if
the model integration times are modest. \citet{drevillon12} describes the
amount of preparation that went into the so-called ``Phase III'' of the search.
Detailed regional atmospheric reanalyses and ocean model hindcasts were
performed to prepare a multi-model high-resolution ensemble of wind and
current fields that were then used to perform a range of backtracking
trajectory integrations.  Similarly, \citet{chen12} included a wind drag
factor and were able to estimate the point of impact for the AF447 accident
based on backtracking the observed debris field.  The method of using a
wind drag coefficient to fine-tune the drift properties was also employed
by \citet{abascal12} to investigate the optimum balance of HF current fields
and wind fields required to backtrack drogued and undrogued drifters.

The 12 articles in this topical collection provide a snapshot more than a
complete overview of the state of object drift modelling and SAR prediction
at sea as it stands today. We hope that by putting together this special
issue we provide a starting point for new workers in the field as well as a
body of references of what has been published earlier. This is particularly
important in an operational field such as SAR planning where a majority of
the work to date is ``grey literature'' in the form of technical reports
that may not be readily accessible or properly vetted through peer review.
SAR planning and object drift modelling demand both mathematical rigour and
experimental finesse to advance further. Peer-reviewed communication is
the most efficient way to achieve this.  It is our hope that this special
issue will contribute to a more academic approach to this exciting field.

\section*{Acknowledgments}
The conference co-chairs would like to express their gratitude to
the organizers and sponsors: IFREMER's Service Hydrodynamique et
Oc\'{e}ano-m\'{e}t\'{e}o, the Norwegian Meteorological Institute, the US
Coast Guard Office of Search and Rescue, JCOMM, Region Bretagne
and the French-Norwegian Foundation. More information about the conference
can be found at
\begin{myindentpar}{1.5cm}
  \underline{\texttt{http://www.ifremer.fr/web-com/sar2011}}
\end{myindentpar}
We are grateful to Springer (publisher of \emph{Ocean Dynamics}) for taking
the topic of SAR into consideration for a special issue. {\O}yvind Breivik is
grateful to The Joint Rescue Coordination Centres of Norway and the Norwegian
Navy for their continued support through funding projects that have allowed
him to help organize these workshops.  The editorial work has also benefited
from the European Union FP7 project MyWave (grant no 284455).  Thanks finally
to Jack Frost, Larry Stone and Henry Richardson for sharing their immense knowledge of the field
of search theory and for helping to unravel the early history of SAR planning.

\bibliography{/home/rd/diob/Doc/TeX/Bibtex/BreivikAbb,/home/rd/diob/Doc/TeX/Bibtex/Breivik} \newpage

\end{document}